\documentclass[twocolumn]{aa}
\usepackage{graphicx}
\usepackage{natbib}
\bibpunct{(}{)}{;}{a}{}{,}
\usepackage{txfonts}
\begin{document}
%
%

   \title{Decay properties of the X-ray afterglows of Gamma-ray bursts}

   \author{B. Gendre
          \inst{1}
      \and
      M. Boer\inst{2}
          }

   \offprints{B. Gendre}

   \institute{Istituto di Astrofisica Spaziale e Fisica Cosmica, Via Fosso del Cavaliere 100, 00133 Roma, Italy\\
              \email{gendre@rm.iasf.cnr.it}
         \and
             Observatoire de Haute-Provence, 04870 St.Michel l'Observatoire, France\\
             \email{Michel.Boer@oamp.fr}
             }

   \date{Received ---; accepted ---}

   \abstract{We present a set of seventeen Gamma-Ray Bursts (GRBs) with known redshifts and X-ray afterglow emission. We apply cosmological corrections in order to compare their fluxes normalized at a redshift of 1. Two classes of GRB can be defined using their X-ray afterglow light curves. We show that the brightest afterglows seem to decay faster than the dimer ones. We also point out evidences for a possible flux limit of the X-ray afterglow depending on the time elapsed since the burst. We try to interpret these observations in the framework of the canonical fireball model of GRB afterglow emission.
   \keywords{Gamma rays:bursts -- X-rays:general}
   }

   \titlerunning{X-ray afterglow decays of GRBs}
   \authorrunning{Gendre \& Boer}
   \maketitle
%

\section{Introduction}
Gamma-ray bursts (GRBs) are among the most enigmatic phenomena in the Universe. Since their discovery in the late 60's \citep{kle73} and until the BeppoSAX revolution, the observer faced a growing sample with no or few constrained properties. One of these, discovered using the the PHEBUS and BATSE observations, was the separation of GRBs in two classes, according to their duration and spectral properties \citep{dezalay96, kou93}. Another step forward was made by the BeppoSAX satellite team, which succeeded in detecting a GRB X-ray afterglow \citep{cos97}. A new generation of automated telescopes such as TAROT \citep{boe99} or ROTSE \citep{ake00} , were built in order to detect quickly the optical counterpart of GRBs.

To date, only long GRB afterglows were observed. Optical observations showed that GRBs are located at cosmological distances \citep{met97}. This allowed to rule out some emission models, and favored the emergence of the fireball model \citep{ree92, mes97, pan98}. In this model a blast wave  propagates into a surrounding medium. The afterglow
emission is described as synchrotron and inverse Compton emission
of high energy electrons accelerated during the shock of an
ultra-relativistic shell with the external medium, while the
prompt emission is due to the internal shocks produced by shells
of different Lorentz factors
within the relativistic blast wave \citep[see ][for a review]{pir99}. In the first development of this model, this medium was uniform, and the blast wave was isotropic. This case is referred as the InterStellar Medium (ISM) model. Two refinements were made later. First, achromatic breaks into the light curve of some afterglows were explained by the non isotropy (i.e. jets) of the blast wave \citep[e.g.][]{pia01}. This model without the isotropy assumption was called the "jet model" \citep{rho97, sar99}. Second, the optical afterglow light curves show in some case bumps associated with type Ic supernova \citep{rei99}. These and X-ray features \citep[e.g. ][]{piro99, ree02} show that long GRBs may be linked with the explosion of a massive star \citep[hypernova; ][]{mes01}. If this is the case, the surrounding medium should not be uniform \citep{che00}: its density decreases with the square of the distance to the central engine, due to the wind  from the progenitor of the GRB. This model is referred as the "wind model" \citep{dai98, mes98, pan98, che99}. Each of these models is defined with a set of parameters, some of them referring to the microphysics ($p$, $\epsilon_e$, $\epsilon_B$), others to the central engine ($E$, $\theta$), and the remaining parameters depending on the surrounding medium ($n$, absorption,...). They also lead to different spectral and temporal characteristics of the afterglows.

Because their brightness allow to detect them at large cosmological distances, and of the growing evidences that at least some long gamma-ray bursts are linked with type Ic supernovae, it is tempting to study whether some of their observational properties may be reproduced from burst to burst. In this case, GRBs could be used as to study the Universe, specifically the region of re-ionization. A first attempt was made by \citet[][ hereafter Paper I]{boe00}: while looking for evidence about optical absorption around the burst, they discovered that  GRB X-ray afterglows with known redshifts have a bimodal luminosity evolution : the faintest GRB afterglows appear to decay slower than the brighter ones. Most of all, they found that bright and faint X-ray afterglows were separated by one order of magnitude in flux one day after the burst. \citet{fra01} showed evidence that the energy released during the prompt emission of the GRB may be roughly constant. In this case the observed range in luminosity is due to beaming effects.

In this paper we reexamine our work presented in Paper I, using a larger sample of X-ray afterglows. At that time our sample was not large enough to give strong conclusions (8 events only). Thanks to the Chandra and XMM-Newton X-ray observatories the sample has grown enough to look for a meaningful correlation. This paper is organized as follow. The data are presented in Sec. \ref{sec_sample}, together with the processing methods. We describe the normalization applied to the GRB afterglow light curves in Sec \ref{sec_theo}. The results are exposed in Sec. \ref{sec_resu} and discussed in Sec. \ref{sec_discu}, before the conclusions.

\section{The data}
\label{sec_sample}

   \begin{table*}
      \caption[]{The GRB sources we used in the present work, together with some data about them. We indicate the satellites used to gather the X-ray data and their redshift, their spectral index and temporal decay. We also indicate the observed flux one day after the burst, the beaming angle of some burst, extracted from \citet{ber03} and some references. We indicate bursts from {\it group I} (see text for definitions) at the top of the table and bursts from {\it group II} at the bottom of the table.}
         \label{table}
     $$
         \begin{tabular}{ccccccccc}
            \hline
            \noalign{\smallskip}
Source & Group &  X-ray    & Redshift & Decay & Spectral & X-ray Flux& Beaming & Reference\\
name   &       & satellite &          & index & index    & @ 1 day & angle   &\\
       &       &           &          &       &          & (10$^{-12}$ erg.cm$^{-2}$.s$^{-1}$) & & \\
            \noalign{\smallskip}
            \hline
            \noalign{\smallskip}
\object{GRB 971214} &I& BeppoSAX          & 3.42      & 1.6 $\pm$ 0.1   & 1.2 $\pm$ 0.4 & 0.23 $\pm$ 0.05 & $>$0.1 & 1, 2\\
\object{GRB 990123} &I& BeppoSAX          & 1.60      & 1.44 $\pm$ 0.11 &1.00 $\pm$ 0.05& 1.8 $\pm$ 0.4   & 0.089& 3\\
\object{GRB 990510} &I& BepooSAX          & 1.619     & 1.4 $\pm$ 0.1   & 1.2 $\pm$ 0.2 & 1.2 $\pm$ 0.2   & 0.054& 4, 5\\
\object{GRB 991216} &I& RXTE, Chandra     & 1.02      & 1.6 $\pm$ 0.1   & 0.8 $\pm$ 0.5 & 5.6 $\pm$ 0.3   & 0.051& 6, 7, 8\\
\object{GRB 000926} &I& BeppoSAX, Chandra & 2.066     & 1.7 $\pm$ 0.5   & 0.7 $\pm$ 0.2 & ---             & 0.14 & 9\\
\object{GRB 010222} &I& BeppoSAX, Chandra & 1.477     & 1.33 $\pm$ 0.04 &1.01 $\pm$ 0.06& 2.7 $\pm$ 0.6   & 0.08& 9\\
\hline
\object{GRB 970228} &II& BeppoSAX         & 0.695     & 1.3 $\pm$ 0.2   & 0.8 $\pm$ 0.3 &0.9 $\pm$ 0.4    & --- & 1, 10\\
\object{GRB 970508} &II& BeppoSAX         & 0.835     & 1.1 $\pm$ 0.1   & 1.1 $\pm$ 0.3 &1.0 $\pm$ 0.4    &0.391& 1, 11, 12\\
\object{GRB 980613} &II& BeppoSAX         & 1.096     & 1.1 $\pm$ 0.2   &    ---        &0.27 $\pm$ 0.07  &$>$0.226& 1, 13\\
\object{GRB 980703} &II& BeppoSAX         & 0.966     & 0.9 $\pm$ 0.2   & 1.8 $\pm$ 0.4 &0.48 $\pm$ 0.07  &0.2& 14\\
\object{GRB 000210} &II& BeppoSAX, Chandra& 0.846     & 1.38 $\pm$ 0.03 & 0.9 $\pm$ 0.2 &0.21 $\pm$ 0.06  & --- & 15\\
\object{GRB 000214} &II& BeppoSAX         & 0.37-0.47 & 0.7 $\pm$ 0.3   & 1.2 $\pm$ 0.5 &0.6 $\pm$ 0.2    & --- & 9\\
\object{GRB 011121} &II& BeppoSAX         & 0.36      & 4$^{+3}_{-2}$   & 2.4 $\pm$ 0.4 &0.6 $\pm$ 0.2    & --- & 9\\
\object{GRB 011211} &II& XMM-Newton       & 2.14      & 1.3 $\pm$ 0.1   & 1.2 $\pm$ 0.1 &0.03 $\pm$ 0.01  & --- &15\\
\object{GRB 030226} &II& Chandra          & 1.98      & 2.7 $\pm$ 1.6   & 0.9 $\pm$ 0.2 &  ---            & --- & 15\\
\object{GRB 030329} &II& RXTE, XMM-Newton & 0.168     &0.9 $\pm$ 0.3    & 0.9 $\pm$ 0.2 &14.3 $\pm$ 2.9   & --- & 15, 16\\
\hline
\object{GRB 980425} &---& BeppoSAX        & 0.0085    & 0.16 $\pm$ 0.04 & 1.0 $\pm$ 0.3 &0.47 $\pm$ 0.07  &---& 17\\
            \noalign{\smallskip}
            \hline
         \end{tabular}
     $$

NOTA : The observations of \object{GRB 000926} and \object{GRB 030226} do not allowed us to extrapolate with meaningfull error bars the light curves. We thus cannot indicate the X-ray flux one day after the burst for these two events. The flux of \object{GRB 000926} was $1.2 \times 10^{-13} \pm 0.1 \times 10^{-13}$ erg.s$^{-1}$.cm$^{-2}$ 2.78 days after the burst and the one of \object{GRB 030226} was $3.5 \times 10^{-14} \pm 0.2 \times 10^{-14}$ erg.s$^{-1}$.cm$^{-2}$ 1.77 days after the burst. The spectral index of \object{GRB 980613} has never been reported, we assumed the value of 1.

references

1 \citet{cos99}

2 \citet{die98}

3 \citet{gal99}

4 \citet{kuu00}

5 \citet{sta99}

6 \citet{pir00}

7 \citet{hal00}

8 \citet{fra00}

9 \citet{pas04}

10 \citet{gal97}

11 \citet{ped98}

12 \citet{piro98}

13 \citet{djo98}

14 \citet{vre99}

15 This work

16 \citet{tie03}

17 \citet{pia00}

   \end{table*}

Our sample is listed in Table \ref{table}. The first column refers to the source name, the third column to the satellites which we used to compute the X-ray afterglow light curves, while the fourth column gives the source redshift. We used only GRBs with known redshift values and exhibiting an X-ray afterglow observed either by at least one of the satellites BeppoSAX, XMM-Newton or Chandra (ACIS imaging mode only). We included \object{GRB 980425} despite its very uncommon properties (very low distance, light curve). For the BeppoSAX bursts we used either published results or data from \citet{pas04}. For the Chandra and XMM-Newton observations we reduced the data using the most up-to-date SAS and CIAO softwares (version 6.0 for the SAS, 3.2 for CIAO). Table \ref{table}
gives the main characteristics of each burst, in the observer frame (no corrections for distance or cosmological effects are included in that table).

We processed the data as follows: we first recalibrate the event lists, using the tasks {\it emchain, epchain} and {\it acis\_process\_events}; these event files are then filtered using the standard procedures depending on the instrument and satellite \citep[we refer the reader to e.g.][for the details of these procedures]{gen03}; we  looked for any period of high background and exclude any occurrence of it. We finally filtered the event files, keeping only events with energy between 0.3 and 10.0 keV. These filtered event lists are used to extract light curves and spectra of each afterglow. We used the XSPEC software \citep{arn96} to fit the spectra with a canonical power law model (taking into account any absorption needed to fit the data), and to derive a mean flux in the 2.0-10.0 keV band. This allows to calibrate the light curves in  flux units.

In the case of \object{GRB 011211}, observed by XMM-Newton, the telescope was re-pointed during the observation. Such a change in the satellite attitude is currently not fully supported by the SAS. We thus discarded the first 5 ks of the observation, and we used only the data recorded after the re-pointing.

We also did not include the second XMM-Newton observation of \object{GRB 030329}, because of the large contamination by a source close to the afterglow position. For this burst we decided to use the results from \citet{tie03}.


\section{The normalization of the X-ray afterglow light curves}
\label{sec_theo}
Because we have decided to compare the flux of several X-ray afterglows of GRBs, we need to normalize them. This normalization is similar to the one computed in Paper I. We correct the fluxes for distance, time dilatation, and energy looses due to the cosmological energy shift. To compute these corrections, we used a flat universe model, with a $\Omega_m$ value of 0.3. 

The distance correction linearly depends on the value of $H_0$, the Hubble constant. Because this value is not well constrained, we normalized the flux to a common distance rather than using the luminosity.  We normalized all X-ray light curves to a common distance corresponding to a redshift of z = 1. These corrections do not take into account any beaming due to a possible jet. The distance used is based on the luminosity distance given by \citet{pen99}.

We corrected the cosmological energy shift as in \citet{lam00}. This implies to know the spectral slope of the afterglow. In order to reduce again uncertainties, we did not correct for the time dilatation effect by interpolating the flux \citep[as in ][]{lam00}; instead, we computed the time of the measurement in the burst rest-frame. This provides a better correction for the time dilatation.

Finally, we restricted the light curves to the 2.0$-$10.0 keV X-ray band, where the absorption is negligible. This allowed us to get rid of any other corrections due to the absorption by the ISM. We point out that the spectral analysis is made with a larger X-ray band in order to reduce the spectral index error bar.

\section{Results}
\label{sec_resu}

\subsection{Results and statistical significance of the groups}

   \begin{figure}
   \centering
   \resizebox{\hsize}{!}{\includegraphics{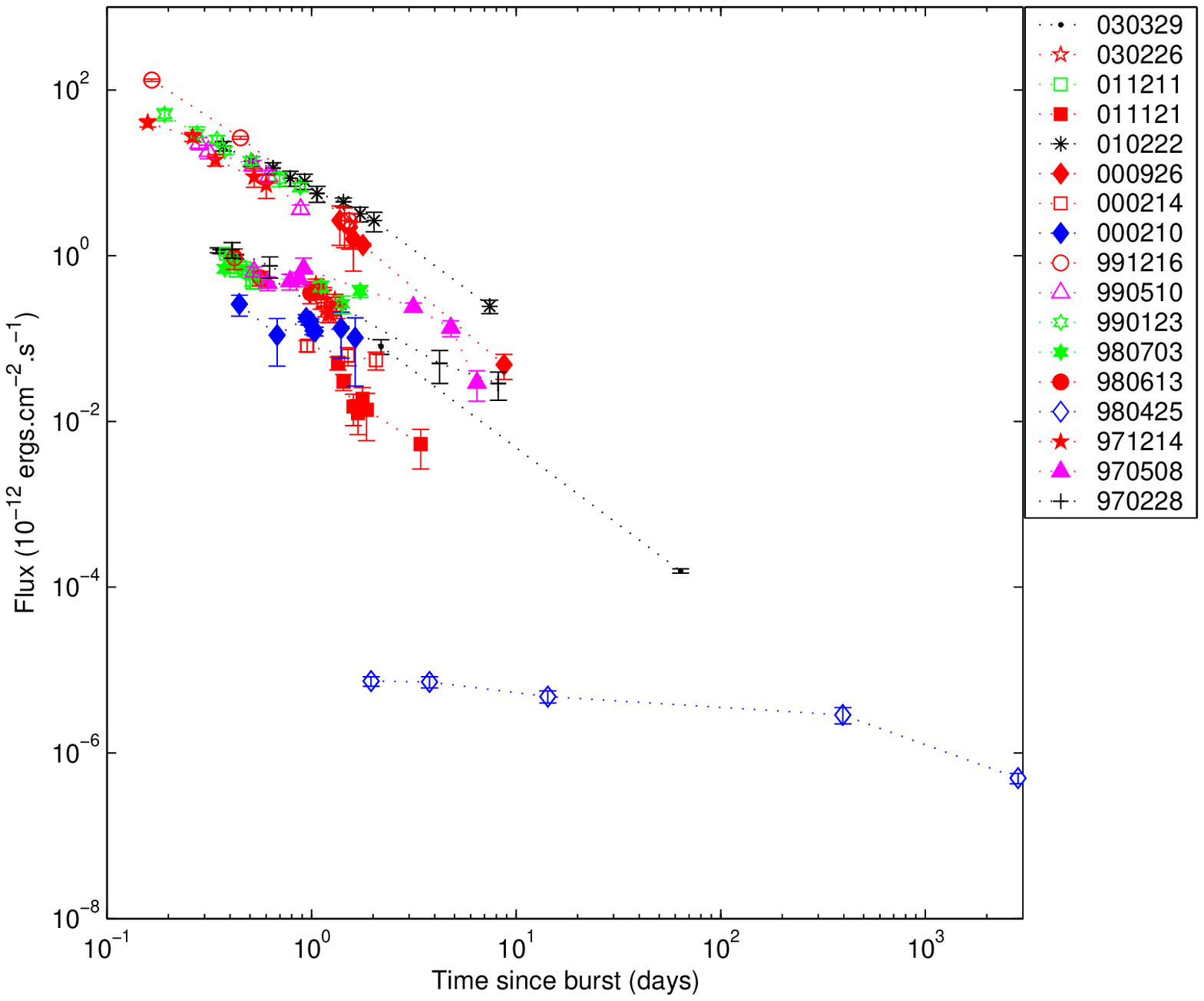}}
   \caption{X-ray light curves of the GRB afterglows rescaled at a common redshift of z=1.}
              \label{Fig1}%
    \end{figure}

\begin{figure*}
   \centering
   \includegraphics{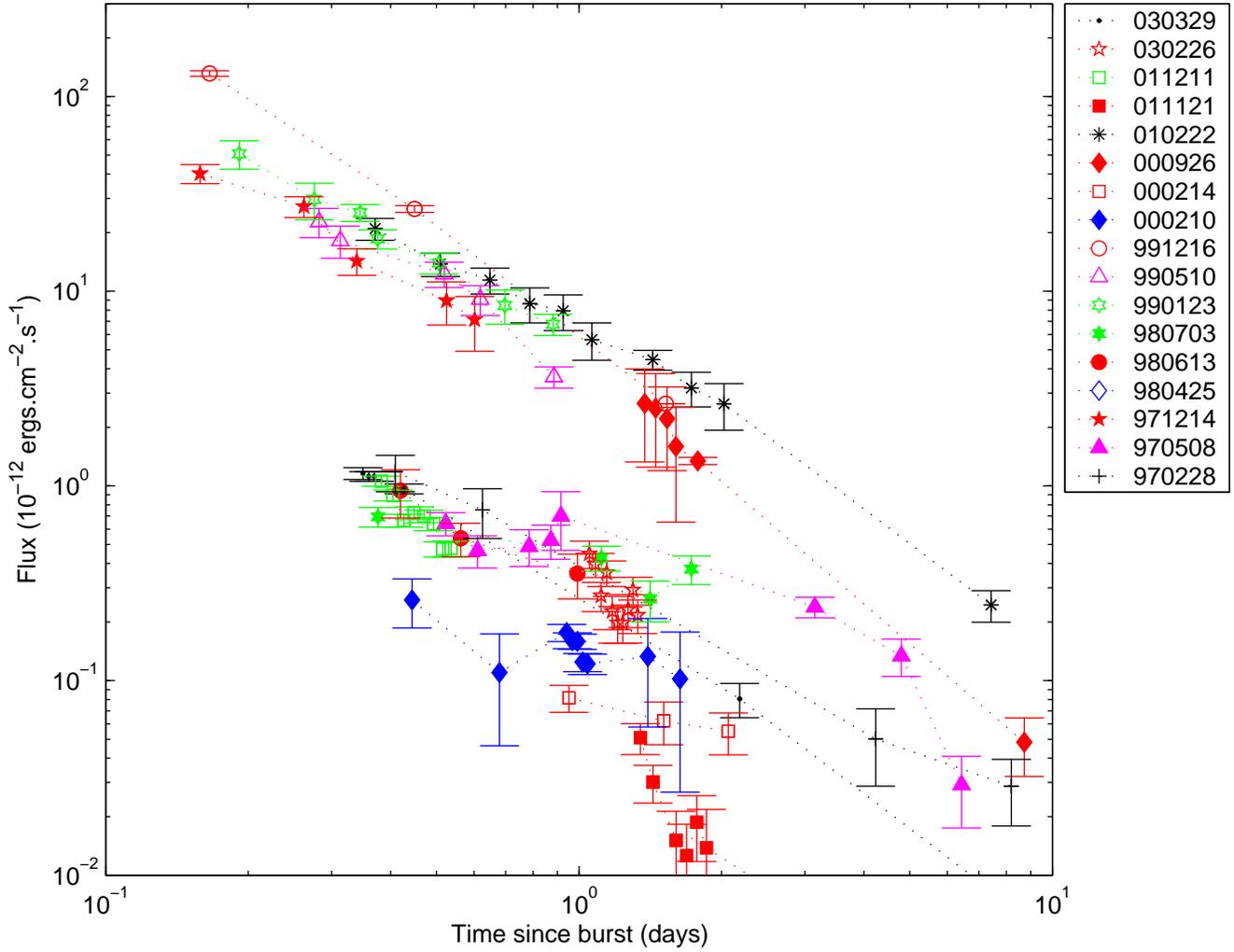}
   \caption{X-ray light curves of the GRB afterglows rescaled at a common redshift z=1. We rescaled the previous figure in order to display the first 10 days of the afterglows.}
              \label{Fig2}%
    \end{figure*}

The result of this comparison is shown in Fig. \ref{Fig1}. A zoom on the clustered data is presented in Fig. \ref{Fig2}. We first note that the two groups noted in Paper I are still present. The sample has now been extended to 17 bursts. All but one lie in one of the two groups. The only (notable) exception is \object{GRB 980425}; however the overall properties of this burst, associated with \object{SN 1998bw} are very peculiar compared to other GRBs \citep[e.g.][]{gal99b}.

We examine now the statistical significance of these two groups. In the following we call {\it group I} the set of GRB afterglows with the brightest luminosity, and {\it group II} the dimmer ones. There are six afterglows in {\it group I}, and ten in {\it group II}. One day after the burst, their mean fluxes are respectively $\sim (5 \pm 2) \times 10^{-12}$ and $\sim (3.5^{+3.5}_{-2.7}) \times 10^{-13} \rm{erg}.\rm{cm}^{-2}.\rm{s}^{-1}$. We tested the hypothesis that a uniform luminosity distribution may cause the observed distribution. The probability to get the observed clustering by chance is only $1.64 \times 10^{-8}$. We computed also the probability to get the observed diagram assuming a power law luminosity distribution, letting the index as a free parameter : the maximum probability is $1.10 \times 10^{-4}$ for an index value of -1. We conclude that the observed clustering in two groups of the actual distribution is significant at at least to the 4 $\sigma$ level.

\subsection{properties of the groups}

We first compute the mean spectral indexes of these two groups: we find $\alpha = 1.0 \pm 0.1$  and $\alpha = 1.2 \pm 0.2$ for {\it group I} and {\it group II} respectively. We note that the spectral indexes are not statistically different.
Since in paper I we indicated that the decay indexes of these two groups were not compatible between them, we computed the mean decay index of the groups. We find $\delta = 1.6 \pm 0.2$ for {\it group I}. If we take into account all bursts of the {\it group II}, we find $\delta = 1.5 \pm 0.9$. However, if we take into account only the bursts with a good decay constraint (hence we ignore GRBs with large error bars, i.e. \object{GRB 011121}, $\delta = 4_{-2}^{+3}$, and \object{GRB 030226}, $\delta = 2.7 \pm 1.6$), we get $\delta = 1.1 \pm 0.2$ (note that we get $\delta = 1.2 \pm 0.2$ if we ignore only \object{GRB 011121}), a result compatible with paper I.
In order to test the statistical significance of the difference between the decay indexes, we computed the probability that the observed repartition is due to only one population of GRBs. Using a Kolmogorov-Smirnov test we find that the probability is 0.13, indicating that this distribution of decay indexes may be due to only one population.

\begin{figure}
   \centering
   \resizebox{\hsize}{!}{\includegraphics{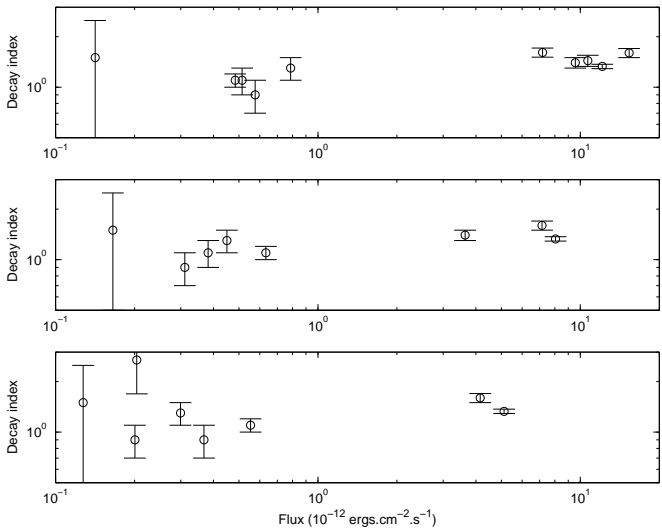}}
   \caption{X-ray decay indexes versus the X-ray flux for several GRBs. The fluxes are taken at 0.6 (top panel), 0.9 (middle panel) and 1.2 (bottom panel) days.}
              \label{Fig5}%
    \end{figure}

Fig. \ref{Fig5} displays the distribution of decay indexes versus the normalized flux at several times. In order to avoid a bias in this repartition, we interpolated the flux value using the two nearest data points, when known. This imply a good time coverage in order to have meaningful repartition. We made this computation at 0.6, 0.9 and 1.2 days. One can see in Fig. \ref{Fig5} the two groups, separated by one order in magnitude at 0.6 days.

\section{Discussion}
\label{sec_discu}

In paper I we tried to explain the nature of these two groups in the framework of the fireball model using the adiabatic/radiative cooling of the fireball. With a value of $p$, the electron power law index, of 2.3, one should observe decay values of 1.68 and 1.22 for {\it groups I} and {\it II} respectively \citep{sar98}: this is compatible with the observed values. However one expects that when the cooling frequency of the fireball is equal to $\nu_m$, the radiative fireball becomes adiabatic \citep{pir99}. This would appears in the light curve as a flattening at that time \citep{sar98}. This is not the case in Fig. \ref{Fig1} : none of the GRBs in {\it group I} displays such a feature. We thus cannot explain the nature of these groups by the adiabatic/radiative cooling of the fireball.

\begin{figure}
   \centering
   \resizebox{\hsize}{!}{\includegraphics{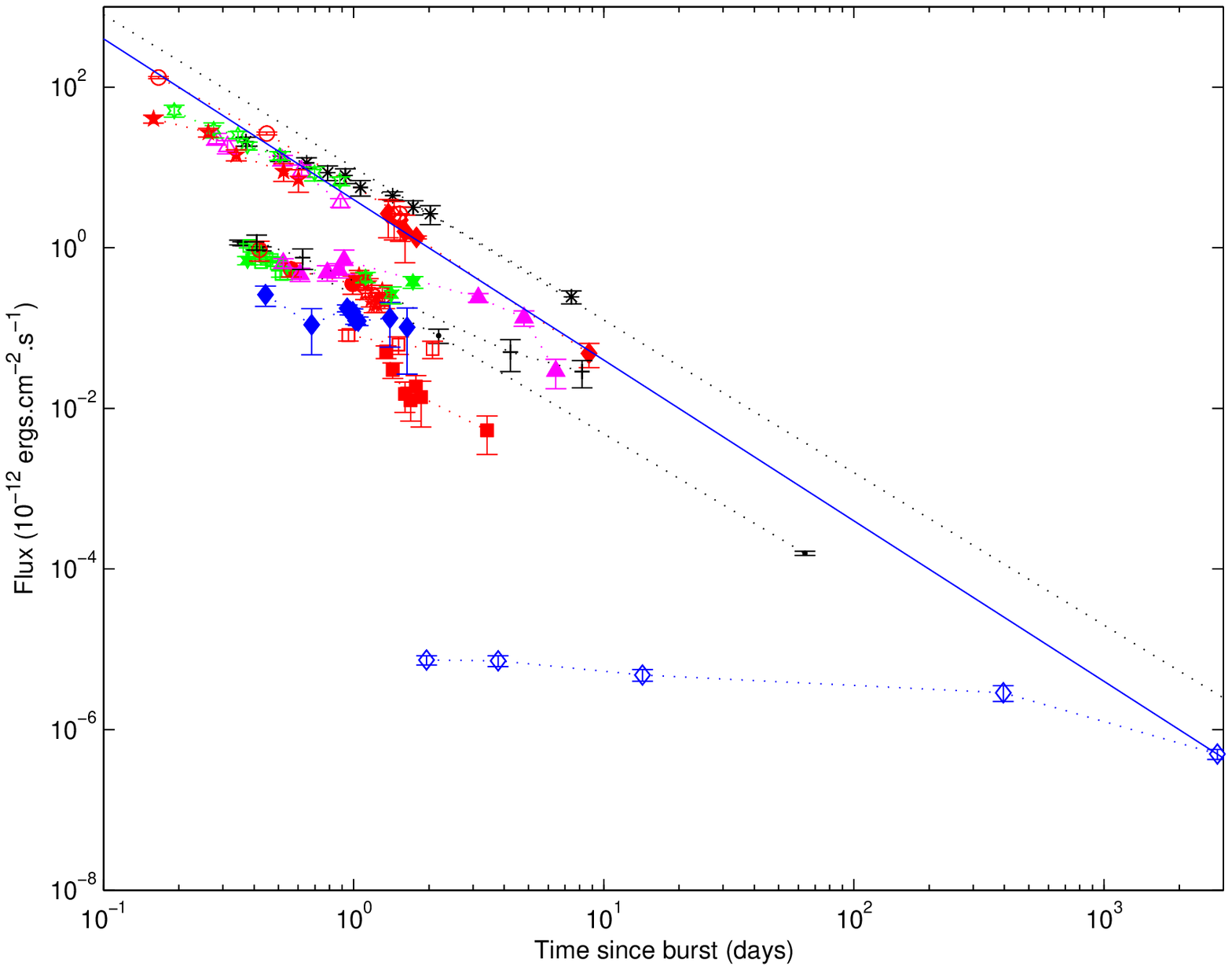}}
   \caption{X-ray light curve of the sample of GRB rescaled at a common distance corresponding to a redshift of z=1. We plotted the limiting fluxes computed as the mean of the bursts from  {\it group I} (solid line) or the as the mean power law corresponding to the brightest burst (dashed line).}
              \label{Fig3}%
    \end{figure}

We used the mean decay index of {\it group I} to interpolate the flux of \object{GRB 971214} at large time (see Fig. \ref{Fig3}). This interpolation seems to define a limit, where there is no data point belonging to {\it group II}. Most of all, some afterglow light curves of the {\it group II} (\object{GRB 970508}, \object{GRB 030329}), which should cross this line, display a steepening. This is also true for the very faint part of this figure : \object{GRB 980425} features a steepening.

One may argue that this limit is not valid, since some bursts of {\it group I} lie above the limit. We computed then a new limit, using the parameters from the brightest burst (see Fig. \ref{Fig3}). In this case we cannot give any firm conclusion about the steepening. We investigate the possibility that this limit is due to instrumental effects : most of the afterglows are observed during the first 10 days of their decay. The only exceptions are \object{GRB 980425} and \object{GRB 030329} which are very very unusual afterglows: We thus cannot use them to stress any model. We conclude that this limit is valid at least for the first ten days of the afterglow.

In order to explain the observed GRB afterglow properties, we first assume that the two groups are not linked together. We use the relationships given in \citet{zha04}. In this framework we suppose that the afterglows from {\it group I} are produced by  a constant density wind from the progenitor, and that afterglows from {\it group II} are expanding in the ISM of constant density. Applying the results from \citet{zha04}, the value of the spectral index should be $(p-1)/2 \sim 0.7$, different from what we observe for {\it group I}.

Another possible explanation could be a difference in the value of $p$ of the two groups. In this case GRBs from {\it group I} have $p = 3.13$ and those from {\it group II} have $p=2.13$, in order to explain all the X-ray data. However, we know from the broad band spectrum fitting of \object{GRB 990123} that $p = 2.28 \pm 0.05$ for this burst \citep{pan02}, not compatible with this hypothesis.

In the following we suppose now that the two groups are linked together. In this case we should observe a break in the low group, otherwise the above mentioned limit would be crossed. Achromatic steepening (also called breaks) in light curves have been associated with jet-like features \citep[see e.g. ][]{tie03}: an increase in the opening angle of the jet imply a steepening in the afterglow light curve \citep{sar99}. We may suppose that the clustering in two groups is due to geometrical effects, such as beaming. Bursts from {\it group I} may have a smaller beaming factor than bursts from {\it group II}.

\citet{ber03} found evidences for a beaming effect on afterglow light curves. We report on Table \ref{table} the beaming angle they derived for our bursts. We clearly see a trend : the bright afterglows have a small beaming angle, as expected in this hypothesis. One should note that this indicates that the fluxes we report here cannot be considered as proportional to the luminosity . This quantity should be obtained by taking into account the beaming factor and the burst parameters (temporal decay and spectral index). This trend shows that the bright (and steepest) afterglows have a small beaming angle. We can then explain the high decay index of {\it group I} by a small beaming angle. {\it Group II} is not affected at early times because of a larger beaming angle.  However, in that hypothesis, the decay index is not compatible with the value expected from the jet model \citep[$\sim 2.3$, ][]{sar99}. In addition, we do not expect such a clustering in two well separated groups from the distribution of GRB beaming angles.

Another possible interpretation is given in the framework of the wind model. In this model, the cooling frequency increase with time \citep{che00}.
As indicated in \citet{che00} the light curve displays a steepening when the cooling frequency pass through the observed frequency: The effect is an increase of the decay index of 0.25. Below the cooling frequency, the decay index is expected to be $\sim 1.6$, compatible with {\it group I}. The index difference between the two groups is 0.5 $\pm$ 0.4, compatible with the model value of 0.25. In this model, if a segregation in decay indexes versus time is expected from different GRBs, connected with the time at which the cooling frequency crosses the X-ray band, there is no particular reason that this occurs at a fixed time for several bursts.

As noted by \citet{che04} some afterglow data can be fitted by either a wind model or an ISM model. On the other hand, \citet{pan01a, pan02} reported that the optical afterglow data cannot be fitted with a wind model for several bursts (e.g. \object{GRB 990123}).

If the GRB source is a very massive star, we expect that the progenitor is surrounded by a medium with a wind profile \citep{che99}. Some refinements have been made later to this model in order to take into account a region where the wind profile ends in the ISM \citep{ram01, che04} with a termination shock. If this is the case, the fireball starts its expansion into a wind profile (the cooling frequency increases with time) then reaches a constant density medium, where the cooling frequency decreases as a function of the time. We can model the light curve as follow (assuming $p = 2.6$) :

\begin{itemize}

\item first, the cooling frequency is below the X-ray, and the flux decay index is $-1.45$,

\item second, the cooling frequency pass through the X-ray band, the light curve displays a steepening and the decay index is $-1.7$,

\item third, the afterglow reaches the termination shock and displays a flattening: The decay index becomes $-1.2$,

\item last, the cooling frequency, which decreases with time as in the ISM model, pass again through the X-ray band, and the light curve displays a steepening and the decay index becomes $-1.45$.
\end{itemize}

The first break is expected to occurs within hours (or less) after the burst, hence it is not observable in our sample. GRB afterglows from {\it group II} can be explained in this framework, since we we observe only the third and last segment mentioned above. Let us suppose now that the cooling frequency does not cross the X-ray band before the afterglow reaches the termination shock. In this case the first and the last segments only will be present, i.e., the light curve will decay as $t^{-1.45}$. This can explain the afterglow light curve of bursts from {\it group I}, i.e. they are observed before or after the termination shock. If this explanation may be valid for the decay indexes values, no constraints are made on the burst brightness.

In this framework we also expect a variation of the spectral indexes between the different segments \citep{zha04}. The spectral index variation should be 0.5. As we observe a variation of 0.2 $\pm$ 0.3 no firm conclusion can be derived.

 Whatever the nature of these groups are, we expect each model parameter to vary within a range. Hence the flux should vary within two extrema, while we observe  clustering around two groups. Variations from this clustering are observed only for two bursts and they can be explained by unusual properties:
\begin{itemize}

\item \object{GRB 011121} deviates from the other bursts of the {\it group II} by a hight decay index. This can be explained noting that the optical light curve of \object{GRB 011121} displayed an achromatic break before the X-ray observation, which may be attributed to a jet feature \citep{gre03}.

\item The case of \object{GRB 000214} is due to the uncertainty on the redshift. All the figures where made with the value of 0.42, while the redshift is badly constrained between a value 0.37 and 0.47.
\end{itemize}

Such a clustering may be the indication that GRB physical parameters (the microphysics parameters, the energy and the beaming angle) and their environmental parameters (density, profile) should lie within a very narrow range, and might even be identical from burst to burst.

\section{Conclusions}

We analyzed the X-ray afterglow data of seventeen GRBs with known redshift values. Our main conclusions are:

\begin{itemize}

\item We confirm our result from paper I \citep{boe00} that GRB X-ray afterglows can be segregated by their decay index and brightness. The interpretation of this clustering within the framework of the standard fireball model remains unclear.

\item We report evidences for a flux limit, i.e. all burst display a flux below $9 \times 10^{-12}$ erg s$^{-1}$ cm$^{-2}$ one day after the burst.

\end{itemize}

Long lasting (1 month or more), and continuous X-ray observations of GRB are needed to confirm the validity of the above reported limit after ten days. The decaying nature of GRB afterglows and their faintness at that times requires the use of very sensitive facilities, such as the XMM-Newton observatory.

We note that it is certainly desirable to increase the statistics, hence the size of the sample. This needs to detect more X-ray and optical afterglows and to determine their redshift, as it will be possible after the launch of the SWIFT experiment. Finally, we note that infrared observations of GRB afterglows, since they are weakly affected by interstellar absorption, may lead to results comparable with those reported here.

\begin{acknowledgements}
We acknowledge the use of the web page of J. Greiner (http://www.mpe.mpg.de/~jcg/grbgen.html). We thank L. Piro, M. DePasquale and A. Galli for useful discussions, comments, and for allowing us to use unpublished data. This work was supported by the EU FP5 RTN 'Gamma ray bursts: an enigma and a tool'. We finally thank P. Meszaros for his comments.

\end{acknowledgements}

\end{document}